\documentclass[traditabstract]{aa} 

\usepackage{graphicx}

\usepackage{txfonts}

\usepackage{rotating} 

\usepackage{natbib}

\bibpunct{(}{)}{;}{a}{}{,} 

\begin{document}

\title{The Hubble sequence: just a vestige of merger events? }


   \author{F. Hammer\inst{1}
 \and
    H. Flores\inst{1}
    \and
 M. Puech\inst{1}
    \and
Y. B. Yang\inst{1,2}
\and
   E. Athanassoula\inst{3}
  \and
   M. Rodrigues\inst{1}
 \and
R. Delgado\inst{1}
         }

   \institute{Laboratoire Galaxies Etoiles Physique et
        Instrumentation, Observatoire de Paris, 5 place Jules Janssen,
        92195 Meudon France\\
              \email{francois.hammer@obspm.fr}
         \and
         National Astronomical Observatories, Chinese Academy of Sciences,
     20A Datun Road, Chaoyang District, Beijing 100012, China
     \and
             Laboratoire d'Astrophysique de Marseille, Observatoire Astronomique de
 Marseille Provence,\\Technopole de l'Etoile - Site de Chateau-Gombert,
  38 rue Fr\'ed\'eric Joliot-Curie, 13388 Marseille C\'edex 13, France
             }

\date{Received March 20, 2009; Accepted: September 2nd, 2009}



\abstract{We investigate whether the Hubble sequence can be reproduced
  by the relics of merger events. We verify that, at
  $z_{median}$=0.65, the abundant population of anomalous starbursts
  -- i.e. with peculiar morphologies and abnormal kinematics -- is
  mainly linked to the local spirals. Their morphologies are dominated
  by young stars and are intimately related to their ionised-gas
  kinematics. We show that both morphologies and kinematics can be reproduced by using gas modelling from Barnes' (2002)  study of major mergers.  Their gas content may be indirectly evaluated by assuming that distant starbursts follow the Kennicutt-Schmidt relation: 
  the median gas fraction is found to be 31\%. Using our modelling to estimate the
  gas-to-stars transformation during a merger, we identify the gas
  fraction in the progenitors to be generally above 50\%.  \newline
All distant and massive starbursts can be distributed along a temporal
sequence from the first passage to the nuclei fusion and then to the
disk rebuilding phase. This later phase has been recently illustrated
for J033245.11-274724.0, a distant compact galaxy dominated by a red,
dust-enshrouded disk. This active production of rebuilt disks is in
excellent agreement with model predictions for gaseous rich
encounters. It confirms that the rebuilding spiral disk scenario -- a strong and
recent reprocessing of most disks by major mergers -- is possibly an important channel for the formation of present-day disks in grand-design spirals. Because half of the present-day spirals had peculiar morphologies and anomalous kinematics at
$z_{median}$=0.65, they could indeed be in major mergers phases 6 Gyrs ago, and almost all at z$\sim$ 1. 

It is time now to study in detail the formation of spiral disks and of
their substructures, including bulge, disks, arms, bars, rings that
may mainly originate from instabilities created during the last major
merger. Many galaxies also show an helicoidal structure, which is probably due to a
central torque, and seems to play an important role in regulating
the angular momentum of the newly-formed disks. }



   \maketitle

%


\section{Introduction}
The tidal torque theory \citep{Peebles76,White} assumes that the
angular momentum of disk galaxies had been acquired by early
interactions. This theory has been supported for a long time: in fact
almost all massive galaxies are regular, 
  including rotational disks and dispersion-supported bulges or their mixes, and
they outline the local Hubble sequence. However if
spiral disks were formed at early epochs - z $>$ 2 - they could have
encountered severe damages from later major interactions. Galaxy collisions
appear to be too frequent, to allow  many disks to survive \citep{Toth92}, and this might happen even at
z$< $1 \citep{Hammer09a}.

How was the Hubble sequence 6 Gyrs ago? Galaxy morphologies strongly
evolve \citep{vBergh2002, vBergh2009, 2004A&A...421..847Z, 2005ApJ...628..160C} towards much
more peculiar structures. This combined with the coeval evolution of
star formation rate and stellar mass densities, of O/H gas abundances
and pair statistics, has prompted us to propose 
the disk rebuilding scenario \citep{Hammer05}. This scenario describes most of these evolutions as due to
a recent merger origin of most spirals.  To reproduce the observed
evolution requires that 50 to 75\% of the present-day spirals have
been formed -- i.e. their disks reprocessed by mergers -- during the
last 8 Gyrs (z$<$ 1). Within this period, observations tell us that
more than half of the stellar mass in spirals has been formed and this
can happen through gas compression occurring during the different
phases of major, gas-rich mergers \citep{Hammer05}.  In this
theory, the disk angular momentum mostly results from the orbital
angular momentum of the last major collision \citep{Puech07, Hammer07, Hopkins09a}.  A more recent epoch for disk formation is
indeed supported by the large decrease with redshift of the fraction of rotationally
supported disks. 
\citet{Neichel08} (hereafter IMAGES-II) found that rotational disks were two times less abundant at
$z_{median}$=0.65, a result that is based on a study combining
detailed-morphology and spatially-resolved kinematics. 

How can mergers be related to the regular local galaxies, of which our
Milky Way has been so often taken as typical? Deep observations 
attest the rather tumultuous history of several nearby galaxies 
that is imprinted in their inner halo (e.g. M31, see \citealt{Brown08,Ibata05}; see also \citealt{2008PASP.120..1145D} for M81). The Milky Way 
appears to be quite exceptional \citep{Hammer07}, possibly 
related to its quiescent merger history. Particularly, its halo 
is the most primordial within galaxies with similar masses 
\citep{2006ApJ...652..277M}  and it shows an angular momentum two times 
smaller than that of typical spirals.

Galaxy simulations can help to test various galaxy formation
scenarios. Assuming large accretions of cold gas flows may reproduce
several correlations, mostly those linking the gas consumption and the
assembly of the stellar mass \citep{Dekel2009}. There is, however,
no convincing observation of significant cold gas-flow in local or distant
galaxies, while mergers are well identified in the local and distant
Universe. \citet{Hopkins09a}  \citep[see also][]{Robertson06}
successfully tested disk survival during merging.  Resulting mergers are producing disks whose
angular momentum differs in direction and amplitude from those of the
progenitors. The predictions of \citet{Hopkins09a} rely on processes
dominated by pure gravitation, which are indirectly affected by
feedback effects. This is certainly true for massive galaxies for
which stellar feedback -including outflows- is unlikely to be
an efficient way to redistribute the material during a collision. If the gas
fraction is sufficient (about 50\%), they predict that the re-formed disk can be the dominant component in the reshaped galaxy.

The rebuilding disk scenario proposes a merger origin for spirals and by extension of the
whole Hubble sequence, from ellipticals \citep{Toomre72, Toomre81, Barnes92} to late type
spirals. In fact, the orbital angular momentum provided by major mergers could solve the angular momentum problem \citep{2002MNRAS.329..423M}.
Considerable work is, however, needed to support the scenario. Could a cosmological distribution of orbital parameters
rebuild small bulges within some rebuilt disks, and large bulges
within others?  Could it reproduce the Hubble sequence statistics of
bulge to disk ratio? \citet{Hopkins09a,Hopkins09b} partially brought a
positive answer to this question. Their model indeed recovers the
well-known correlation between bulge-to-disk ratio and mass.

Observations may prove or invalidate the rebuilding disk scenario. It
is well known that the gaseous content of galaxies increases rapidly
with redshift. But does it reach the values required to rebuild a disk
in case of mergers? We must also observe the details of the physical
processes in galaxies at different epochs and examine directly their
evolution. At very high redshift (z$>$ 2) cosmological dimming
prohibits the examination of the optical radius of a disk even with
the largest space telescopes, and it is difficult yet to gather a
representative sample of such galaxies. At intermediate redshifts the
situation is much better even if it needs pushing the present
observational tools near their limits. Up to z=0.4 and z=1.3 the
optical radius of a redshifted Milky Way can be retrieved
from GOODS and UDF imageries, respectively.

The IMAGES study aims to identify the physical
processes that link distant (z $\sim$ 0.65) to local galaxies.  Its selection is
limited by absolute J-band magnitude ($M_J$(AB)$\le$ -20.3), a quantity relatively well
linked to the mass \citep[][hereafter IMAGES-I]{Yang08}. Using such a limit, \cite{Delgado09} have shown that z $\sim$ 0.65 galaxies have to be the progenitors of local galaxies selected in a similar way. They find that the fraction of E/S0 did not evolve since the last 6 Gyrs, while spiral galaxies were 2.3 times less abundant. They indeed use a quite restrictive method to classify morphologies, assuming that spiral galaxies in the past should have similar properties than what they possess today. Using such a morphological classification, IMAGES-II \citep{Neichel08} demonstrated an excellent agreement between morphological and kinematical classifications. In other words, most rotating galaxies (80\%) show spiral morphologies while most galaxies (90\%) with anomalous kinematics present peculiar morphologies. The above results have an important impact: anomalous kinematics of the gaseous component (from the ionised gas, $[OII]$$\lambda$3726;3729 lines) is almost always linked to anomalous morphological distribution of the stars. Altogether the above results imply that {\it more than half of the present-day spirals} had anomalous kinematics and morphologies, 6 Gyrs ago.

Anomalous galaxies are also responsible \citep{Flores06} of the most striking evolution of the Tully-Fisher relation, i.e. it is heavily scattered at z$\sim$0.6-1 \citep{2005ApJ...628..160C}. Major mergers can reproduce this evolution \citep{Covington09} as well as a similar trend for the
$j_{disk}$-$V_{flat}$ relationship \citep{Puech07}. The goal of this paper is to verify whether the observed evolution can be mostly related to merger events, i.e. to test if the rebuilding disk scenario is consistent with the observed evolution of morphology and kinematics. We thus defined 3 different morpho-kinematical classes following the Table 4 of  \citet{Neichel08}:\newline
- rotating spiral disks are galaxies possessing a rotating velocity field, including a dispersion peak at the dynamical centre \citep[see e.g. ][]{Flores06}, and showing the appearance of a spiral galaxy;\newline 
- non-relaxed systems are galaxies that have a strong discrepancy from
a rotational velocity field and whose morphology is peculiar;\newline 
- and semi-relaxed systems that possess either a rotational velocity field and a peculiar morphology or a velocity field discrepant from a rotation and a spiral morphology.\newline
Table 1 summarises the statistics at $z_{median}$=0.65 and compares them to local galaxies from SDSS \citep{Nakamura2004}. Note that similar statistics combining kinematics and morphology does not exist for both local galaxies and quiescent distant galaxies, and thus the corresponding fraction of non-relaxed systems is still inaccurate. Indeed \cite{Delgado09} found that 10\% and 25\% of local and quiescent distant galaxies show peculiar morphologies, respectively. Given this, Table 1 provides only a lower limit of the fraction of distant galaxies that show anomalous properties and that are the progenitors of present-day spirals.


 \begin{table}
\caption{Morpho-kinematical classification of 52  $z_{median}$=0.65 galaxies from \citet{Neichel08} (see their Table 4); for comparison, the last column shows the fractions derived from the SDSS \citep{Nakamura2004} for galaxies in the same mass range (e.g. Hammer et al., 2005). CFRS \citep{Hammer97} found that 60\% of $z_{median}$=0.65 galaxies have spectra with $W_0(OII)\ge$15\AA~ and are classified as starbursts. References are \citet{Neichel08}: N08, \citet{2004A&A...421..847Z}: Z06, \citet{Nakamura2004}: N04 and \citet{Hammer05}: H05.}
{\scriptsize
\begin{tabular}{lcccc} \hline
Redshift   & $z_{median}$=0.65 & $z_{median}$=0.65 & $z_{median}$=0.65   & z=0 \\ 
   & Starburst (60\%) & Quiescent (40\%) & All   & local \\ 
       & $W_0(OII)\ge$15\AA~ & $W_0(OII)<$15\AA~ & All & All \\
References       & N08  & Z06  & N08 & N04 \\
       & &  & & H05 \\ 
Type       &  &    &   & \\ \hline
 & & & & \\
 E/S0        &   0\% &  57\% & 23\% & 27\%\\
Rotating spiral disks   &  23\% & 43\%  &  31\% & 70\%\\
Non-relaxed \&  &  77\% & 0\% &  46\% & $\sim$ 3\%\\
intermediate systems  &  &    &   & \\ 
 & & & & \\\hline
Galaxies with &  &    &   & \\ 
anomalous &  68\% & 0\%   & 41\%  & \\ 
kinematics &  &    &   & \\ \hline
\end{tabular}
}
\end{table}

Morphological and kinematics properties are coming from the
IMAGES survey and the complete picture has been also provided by deep photometric and
spectroscopic measurements necessary to estimate their SFR \citep[both in UV
and IR, see also ][]{Puech09a}, their stellar masses and their
chemical and stellar population decomposition.  
Section 2  describes our procedure to test whether or not morpho-kinematical properties of distant starbursts can be reproduced by galaxy interactions processes, including during the remnant phase. In section 3 we show the overall properties of distant starbursts, including their gas richness that is of crucial relevance to infer whether mergers may lead to disk rebuilding.  In section 4, we discuss the results and conclude in section 5 on the validity of the disk rebuilding scenario. 
 Throughout the
paper, we adopt $H_0=70$ km/s/Mpc, $\Omega _M=0.3$, $\Omega
_\Lambda=0.7$ and the $AB$ magnitude system.

\section{Could distant starbursts properties be reproduced by merger or merger remnants? }

\subsection{Detailed analyses of individual distant galaxies }

 Detailed analyses of four distant galaxies of the IMAGES study have
 been performed by \citet{Puech07,Puech09a}, \citet{Hammer09b} and
 \citet{2008arXiv0812.1593P}, and four other studies of individual
 galaxies are in progress (Yang et al., 2009; Fuentes et al. and
 Peirani et al. in preparation). By modelling gas motions as well as
 morphologies, these studies have shown their ability in reproducing
 the properties of distant galaxies with a similar accuracy to what is
 done for nearby galaxies. \citet{Puech07} have demonstrated that
 spatially resolved kinematics is sufficiently sensitive to detect the
 infall of a 1:18 satellite in a z=0.667 galaxy.
 \citet{2008arXiv0812.1593P} identified a giant and starbursting bar
 induced by a 3:1 merger, and simulated both morphologies and the
 off-centre dynamical axis. In this case, the gas pressured in the
 tidally formed bar has condensed into young and blue stars.
 \citet{Hammer09b} identified a compact LIRG dominated by a
 dust-enshrouded compact disk that surrounds a blue, centred helix
 (so-called a "two arms-plus-bar" structure). They interpret (see
 their Fig. 7) this structure as regulating the exchanges of the
 angular momentum and possibly stabilising the new disk
 \citep{Hopkins09a}. Indeed gas inflows along an helix are usual in
 simulations of mergers, especially in inclined and polar orbits. This
 gaseous-rich galaxy appears to be an archetype of a disk rebuilding
 after a 1:1 or a 3:1 merger with an inclined orbit. Puech et al.
 (2009a) demonstrated that the presence of ionised gas without stars
 near a highly asymmetric disk can be only reproduced by a remnant of
 a merger.

 These studies have been successful because they compared simulations
 of the gas phases to observations of both the morphology and the
 ionised gas motions. Morphologies of starbursts -especially the
 numerous blue or dusty regions- are mostly relics of gas phases
 recently transformed into young stars that ionise the gas. Thus a
 common physical mechanism should reproduce them together with the
 observed large-scale motions of the ionised gas. Within most
 starbursts, the light is indeed dominated by $\le$ 100Myrs-old stars
 and at large distances, spatially-resolved kinematics only detect
 large-scale motions, with typical scales of $\sim$3kpc. A typical
 motion of 100km/s would cross such a length scale during $\sim$
 50Myrs (32 Myrs for motions parallel to the sky plane). Thus many
 morphological features with blue colors \citep[bars, rings and
   helixes, see ][]{2008arXiv0812.1593P,Hammer09b} should be imprints
 of the gas hydrodynamics and they can be compared to the gas
 kinematics.

\subsection{A general method to compare galaxy-simulations to distant starbursts}

For reasons of homogeneity, we study here the sub-sample of 33 IMAGES
starbursts (see IMAGES-I) observed in the CDFS-GOODS. This sub-sample
is representative of $M_J$(AB)$\le$ -20.3 starbursts (see IMAGES-I).
Two galaxies have been rejected from the original sample of IMAGES-I,
one (J033210.76--274234.6) because it turns out not to be a starburst
\citep{Yang09} and another one (J033250.24-274538.9) because the
HST/ACS images are corrupted. We have verified that this sub-sample is
representative of the stellar-mass and star formation densities at
$z_{median}$=0.65 \citep[see e.g.][]{Ravi07}. In this sample we do
find only 6 rotating spiral disks to which we add one giant spiral
(J033226.23-274222.8 that is also rotating while it likely experiences
a satellite infall causing a small shift in the observed dispersion
map \citep{Puech07b}. Note also that one of the rotating spiral galaxy
(see Fig. 1, right) is within a confirmed interaction with an
elliptical galaxy. The 26 other galaxies all show peculiar
morphologies and/or anomalous kinematics and are classified as such as
non or semi-relaxed systems.

\begin{figure}
   \centering
\includegraphics[width=9cm]{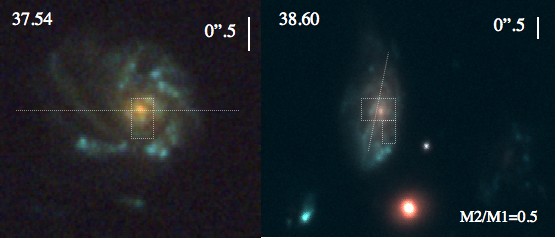}
   \caption{b+v, i and z combined images of 2 rotating disks identified by IMAGES. Dotted line is the superposition of the dynamical axis, dotted squares indicates the dispersion peak element. On the left, the dynamical and optical  axes are aligned, and the dispersion peak is at the mass center, as expected for a rotation \citep[see][]{Flores06}. On the right (J033238.60-274631.4) there is a slight misalignment of the dispersion peak that is likely caused by the nearby passage - 15 kpc- of a bulge-dominated galaxy, causing the observed burst of star formation at the bottom edge of the spiral galaxy. The velocity difference between the two galaxies is 540 km/s, a value based on spectroscopy. }

              \label{Fig1}
    \end{figure}

It is an Herculean task to analyse in details all the considerable
amount of data for each of these galaxies, as it has been described
for few galaxies in section 2.1. The accurate modelling of both
morphology and kinematics takes several months, from two to six months
for a well-experimented user. This is due to the wide complexity of
the morphologies and kinematics in these non-relaxed galaxies as well
as the large parameter space offered by the simulations (mass ratio,
orbit, temporal phase, peri-centre radius and parameters of the
encounters, viewing angles). Our goal here is restricted to the
following question: \emph{ Can we sketch both the morphological and
kinematics main properties of the 26 non-(or semi-)relaxed galaxies by
merger or merger remnants?}

Recently, \citet{2009AJ....137.3071B} have defined a modelling tool to
identify merger orbital parameters. It allows to change many
parameters including the viewing angle. However at high-z we cannot
identify low surface brightness tidal features. We propose here to
adapt a similar modelling tool for high-z observations, also allowing
changes of the viewing angle. We then used the models from
\citet{Barnes02}, that include 12 configurations with mass ratio
ranging from 1:1 to 3:1, orbits from INClined, DIRect, POLar and
RETrograd and pericenter radii from $r_p$=0.2 to 0.4 \citep[see][for
  more details]{Barnes02}. We have recovered the ZENO code
source\footnote{The ZENO simulation code was retrieved from the Josh
  Barnes website (http://ifa.hawaii.edu/~barnes/software.html). The
  code has been improved by the intensive utilization of GPU
  (execution 10 times faster than CPU). } and follow all the
parameters in \citet{Barnes02}, except for few differences for the
values of the pericenter radius. However, the adopted values in our
simulations make them resembling the Barnes' videos. In some cases we
had to invert the spin of interacting galaxies to match the observed
velocity gradient. The number of particles in each simulation is 95040
slightly larger than in \citet{Barnes02}.

 We then developed the interactive system based on Barnes' command
 "snapview" which allows us to display/rotate simulation in 3D space.
 We thus improved it by generating the projected image (morphology)
 and velocities from which we can mimic the IFU observation. After
 matching the morphology, the velocity of each particle is projected
 to line-of-sight direction. Then we mimicked the observation of an
 IFU by calculating the mean velocity and velocity dispersion for
 those particles that are projected into an IFU pixel. We have first
 tried to reproduce the gross morphological features and then we have
 tried to catch the kinematics by dithering the IFU grid within half
 IFU pixel, and by rotating the system within $\sim$ 5-10 degrees.
 Figure 2 displays the final result after rotation for morphology,
 velocity field and dispersion maps which compare the best to the
 observations within the adopted grid of models.

To measure the quality of the model, three sets of parameters are
  considered: the ones from morphology, from velocity field and from dispersion
  maps, respectively. Each set of parameters is graded from 0 (failure) to 2 (good fit). For each of them
  the following criteria have been considered: 
\begin{itemize}
 \item Morphological parameters: only large scale morphological
   structures were considered, including the presence of multiple
   nuclei, bars, ring, arms, helix etc.;
 \item  Velocity field parameters: orientation of the main velocity gradient(s), but not 
 their amplitude(s);                                   
 \item  Dispersion map parameters: position of the minima and maxima, but not
 their amplitudes.
\end{itemize}
The quality of a model is given by the final grade which is the sum of
the grades for each three sets of parameters. Three of us have
independently classified each selected model (LA, FH and MP).
Individual grades has been then compared after the completion of the
whole classification. An excellent agreement have been found between
the individual classifications, with only three major disagreements
within 26 objects. The final grade is the median value and has been
provided after a face-to-face meeting between the three classifiers.
Table 2 provides the individual grades as well as the resulting final
one. We assume in the following that a robust model has been obtained
when the final grade is equal or above 4, while median grade of each
three sets of parameters has to be above or equal to 1. This means
that we are able to reproduce simultaneously morphology and kinematics
in a quite robust way. We then find that 17 starbursts among 26 are
robustly reproduced by our simple modelling method (see Table 2 and Figure
2).

\begin{table}
\caption{Classification of the quality of the model in reproducing morphology and kinematics.}
\begin{tabular}{lcccccccc}\hline
IAU name             &Class$^a$& \multicolumn{2}{c}{Cl$_{LA}^b$}     &   
\multicolumn{2}{c}{Cl$_{FH}^b$}  &   \multicolumn{2}{c}{Cl$_{MP}^b$}   & 
FC$^c$ \\\hline
J033210.25-274819.5 & NR& 212&  5 &  222& 6& 222 &  6 & 6\\        
J033212.39-274353.6 & RD& - &  - &  - & -& - &  - & 6\\        
J033213.06-274204.8 & SR& 221&  5 &  122& 5& 211 &  4 & 5\\       
J033214.97-275005.5 & NR& 101&  2 &  111& 3& 111 &  3 & 3\\        
J033217.62-274257.4 & NR& 022&  4 &  212& 5& 122 &  5 & 5\\          
J033219.32-274514.0 & NR& 112&  4 &  202& 4& 112 &  4 & 4\\         
J033219.61-274831.0 & NR& 121&  4 &  121& 4& 121 &  4 & 4\\
J033219.68-275023.6 & RD& -&  - &  -& -& - &  - & 6\\
J033220.48-275143.9 & NR& 111&  3 &  110& 2& 010 &  1 & 2\\
J033224.60-274428.1 & NR& 222&  6 &  122& 5& 222 &  6 & 6\\
J033225.26-274524.0 & NR& 120&  3 &  111& 3& 110 &  2 & 3\\
J033226.23-274222.8 & SR& - &  - &  - & -& - &  - & 6\\
J033227.07-274404.7 & NR& 222&  6 &  122& 5& 222 &  6 & 6\\
J033228.48-274826.6 & NR& 111&  3 &  120& 3& 220 &  4 & 3\\
J033230.43-275304.0 & NR& 122&  5 &  211& 5& 212 &  5 & 5\\
J033230.57-274518.2 & NR& 010&  1 &  101& 2& 100 &  1 & 1\\
J033230.78-275455.0 & RD& - &  - &  - & -& - &  - & 6\\
J033231.58-274121.6 & RD& - &  - &  - & -& - &  - & 6\\
J033232.96-274106.8 & NR& 121&  4 &  122& 5& 221 &  5 & 5\\
J033233.90-274237.9 & NR& 212&  5 &  221& 5& 222 &  6 & 5\\
J033234.04-275009.7 & SR& 211&  4 &  121& 4& 121 &  4 & 4\\
J033234.12-273953.5 & NR& 010&  1 &  120& 3& 020 &  2 & 2\\
J033237.54-274838.9 & RD& - &  - &  - & -& - &  - & 6\\ 
J033238.60-274631.4 & RD& - &  - &  - & -& - &  - & 6\\
J033239.04-274132.4 & NR& 121&  4 &  120& 3& 120 &  3 & 3\\
J033239.72-275154.7 & NR& 220&  4 &  220& 4& 220 &  4 & 4\\
J033240.04-274418.6 & NR& 220&  4 &  220& 4& 220 &  4 & 4\\
J033241.88-274853.9 & SR& 222&  6 &  222& 6& 222 &  6 & 6\\
J033244.20-274733.5 & NR& 210&  3 &  100& 1& 100 &  1 & 1\\
J033245.11-274724.0 & SR& 222&  6 &  222& 6& 222 &  6 & 6\\
J033248.28-275028.9 & SR& 221&  5 &  221& 5& 221 &  5 & 5\\
J033249.53-274630.0 & NR& 201&  3 &  100& 1& 001 &  1 & 1\\
J033250.53-274800.7 & NR& 222&  6 &  122& 5& 122 &  5 & 5\\\hline
\end{tabular}
Notes: \\
$^a$ Morpho-kinematical classification (RD: rotating spiral disks, NR: non relaxed systems and SR: semi-relaxed systems, see section 1).\\
$^b$ Quality grade of the model for the three sets of parameters (morphology, velocity field and the $\sigma$-map). Each one is graded from 2 (good fit) to 0 (failure). The last grade is the sum of the three previous ones.  \\
$^c$ The final classification grade is the median value of the three grades attributed by each individual and ranges from 6 (excellent agreement) to 4 (robust fit) and then to 3 and below (not reliable fit).   \\

\end{table}

Let us now examine whether or not these robust fits gives convincing
credit to the merging hypothesis. The main limitation of this exercise
is obviously given by the discrete number of orbits and of mass ratio
between the modelled encounters. Since we allow to invert the spin of
one or two of the interacting galaxies, the total number of models is
48. Notice that we had to invert the spin for 5 cases among the 17
robustly modeled starbursts. In fact our initial motivation to choose
such a methodology, besides the obvious question of time consuming,
was coming from our initial experience. For each individual detailed
modelling (see section 2.1) our initial guess for the orbit and mass
ratio had been close to the final result, or in other words, only few
changes of the latter quantities have allowed to recover the
amplitudes of the kinematical parameters as well as most of the
morphological details. It is indeed possible that the choice of the
orbits by \citet{Barnes02} is quite representative of a cosmological
distribution of merger orbits. Besides this, the use of 1:1 and 3:1
mass ratio might be also sufficient to reproduce the gross features of
many major mergers.

Our models, including the robust ones, are not unique and this
probably applies also to detailed models that have been discussed in
section 2.1. The important question to address here is the possible
degeneracy of the methodology applied here, that would be the case if
the number of model parameters is significantly larger than the number
of constraints. To test this we propose to consider two robust models,
one of an on-going merger with two identified nuclei and one of a
merger remnant. J033210.25-274819.5 is a galaxy showing two components
with a strong colour difference (see Fig. 2) of 1.5 magnitude in
observed $(b-z)_{AB}$ \citep[see ][]{Neichel08}. To test the merger
hypothesis we simply assumed that the two components are the relics of
a 3:1 mass ratio merger with a DIRect orbit just before the first
passage. This galaxy shows two peculiarities in its kinematics. First
the dynamical axis is seriously offset from the main optical axis and
it points towards the secondary blue component and second, the
dispersion peak is also offset in the direction of the secondary
component. The constraints that are needed to be reproduced
are:\newline - the location of the secondary component (2) and the
presence of a small bar (1) in the center of the main
component;\newline - the main dynamical axis (2) that is shifted
towards the secondary component as well as the presence of a secondary
dynamical axis (1) that follows the main optical axis of the main
component;\newline - the location of the dispersion peak (2) as well
as the minima (1 to 3) in the dispersion maps. \newline The parameters
required to reproduce these $\ge$ 9 constraints include only the mass
ratio, the elapsed time during the merger and the orbit, i.e. a much
smaller number of parameters when compared to the number of
constraints. This small number of parameters is due to the fact that
many parameters are fixed including the parabolic orbit, gas fraction
and pericenter radius (as it is the case for 3:1 merger in
\citet{Barnes02}). Other parameters such as the total mass, the
baryonic fraction, the profile of dark and baryonic matter have been
also fixed during the simulation, consistently with the fact that we
are not reproducing amplitudes of the kinematical properties. The
second galaxy is J033232.96-274106.8 which is a compact galaxy. Fig. 2
shows not only the morphology and the kinematics but also the residual
image after having removed the best-fitted luminosity profile that is
a n=1 Sersic index with a 6.4 HST/ACS-pixel disk radius. The residual
shows a so-called helix structure that is reproduced by the simulation
(gas component) which also reproduces the dynamical axis, the
structure of the velocity field as well as the dispersion peak and
most of the minima. As for the former example, in this (a 3:1 POLar)
merger the number of free parameters is very small (3) and far below
the number of constraints (8) to reproduce.

Another degeneracy that might affect the modeling process is the one
associated with the uniqueness of the best model itself. Indeed, even
if the number of constrains exceed the number of model parameters as
discussed above, there remains the possibility that at constant number
of free parameters, several different models could match the
observations with a similar quality. This model-degeneracy was
extensively discussed by \citet{2009AJ....137.3071B}. They pointed out
that such a degeneracy can be broken by identifying specific features
in the phase space (ie, the 6D space of positions and velocities)
after the encounter, because such features (e.g., tidal tails) allow
us to trace back the initial configurations of the progenitors. They
showed that such a methodology allowed them to robustly constrain "the
disk orientations, viewing angles, time since pericenter, pericentric
separation, and scale factors", while they did not examine "errors in
center-of-mass position and velocity". In spirit, our approach is very
similar: we examined the 6D phase space using projections of different
moments of the phase function (morphology, velocity field, and velocity
dispersion maps) and tried to reproduce specific signatures, which are
listed above. Hence, we are quite confident that the methodology used
in this paper allows us to constrain efficiently the merging phases as
well as the disk inclinations provided that the time since pericenter
does not correspond to the latest merging phases where such specific
features tend to vanish. But strictly speaking, it is clear that it
remains difficult, and probably even impossible, to claim that these
models are truly unique, something which is anyway inherent to any
modeling work, whatever the adopted methodology and/or data quality
are.

In the sample of 33 emission line galaxies there are six rotating
spiral galaxies, five of them being isolated and one being in
interaction with an elliptical galaxy (see Fig. 1). Another galaxy is
almost similar to a rotating disk although it experiences a satellite
infall. Among the other 26 galaxies, 17 of them have their
morphologies and kinematics robustly reproduced by a merger model with
a number of parameters that is far below the number of constraints
provided by the observations. There are nine other galaxies for which
our modelling as mergers appears less secure. All these galaxies have
anomalous velocity fields and peculiar morphologies and generally
their dynamical axes show significant offset with to the main optical
axes. These galaxies have thus similar properties than those of the
robustly modelled mergers or remnants discussed above. Some of them
are obviously in strong interaction (e.g. J033220.48-275143.9) or are
very likely merger remnants (e.g. J033214.97-275005.5 and
J033230.57-274518.2) from their extremely distorted morphologies and
kinematics. The larger uncertainty in modelling them could be due to
the limitations of the templates used here. We classify them as
possible mergers or merger remnants in the following.

\setcounter{figure}{1}
\begin{figure*}
\begin{tabular}{l}
 IAU name and associated model \\
\includegraphics[height=2.9cm]{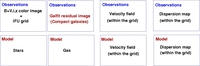} \\
\end{tabular}

\begin{tabular}{ll}
J033210.25-274819.5 (DIR 3:1 $r_{peri}$=0.2) &  J033213.06-274204.8  (DIR 3:1 $r_{peri}$=0.2) \\
\includegraphics[height=2.9cm]{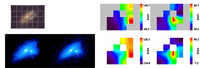} & \includegraphics[height=2.9cm]{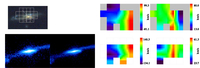} \\
J033214.97-275005.5  (INC 1:1 $r_{peri}$=0.2) &  J033217.62-274257.4  (INC 1:1 $r_{peri}$=0.4)  {\it Spin inverted for both}\\
\includegraphics[height=2.9cm]{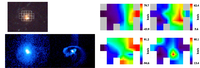} & \includegraphics[height=2.9cm]{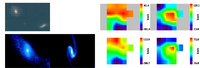} \\
J033219.32-274514.0  (INC 1:1 $r_{\rm peri}$=0.4){\it Spin inv. for both} &J033219.61-274831.0  (POL 3:1 $r_{\rm peri}$=0.2)\\
\includegraphics[height=2.9cm]{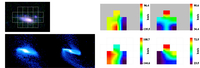} & \includegraphics[height=2.9cm]{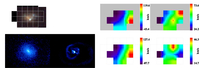} \\
 J033220.48-275143.9  (POL 3:1 $r_{\rm peri}$=0.2)  {\it Spin inv. for the small galaxy} & J033224.60-274428.1  (INC 3:1 $r_{\rm peri}$=0.2)\\ 
\includegraphics[height=2.9cm]{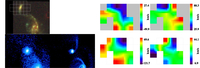} & \includegraphics[height=2.9cm]{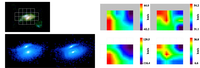} \\
 J033225.26-274524.0  (INC 3:1 $r_{\rm peri}$=0.2)&  J033227.07-274404.7  (INC 3:1 $r_{\rm peri}$=0.2) \\
\includegraphics[height=2.9cm]{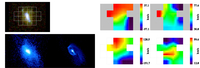} & \includegraphics[height=2.9cm]{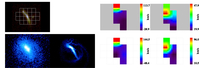} \\
 J033228.48-274826.6  (POL 1:1 $r_{\rm peri}$=0.2)& J033230.43-275304.0  (POL 3:1 $r_{\rm peri}$=0.2) \\
\includegraphics[height=2.9cm]{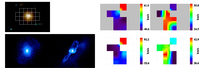} & \includegraphics[height=2.9cm]{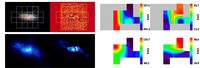}   \\
\end{tabular}
\caption{Comparison between observations and models for the 26 distant starbursts that show non or semi-relaxed properties from their morphologies and kinematics. The top sketchy boxes indicates the signification of each panels. See Table 2 for the classification of the modelling and Table 3 for the general properties of these galaxies.}
\end{figure*} 

\setcounter{figure}{1}
\begin{figure*}
\begin{tabular}{ll}
J033230.57-274518.2  (POL 1:1 $r_{\rm peri}$=0.4) &J033232.96-274106.8  (POL 3:1 $r_{\rm peri}$=0.2) \\
\includegraphics[height=2.9cm]{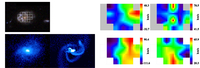}&  \includegraphics[height=2.9cm]{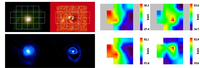}\\
 J033233.90-274237.9  (RET 1:1 $r_{\rm peri}$=0.4)&  J033234.04-275009.7  (RET 1:1 $r_{\rm peri}$=0.2) \\
\includegraphics[height=2.9cm]{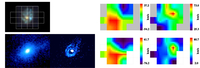}&  \includegraphics[height=2.9cm]{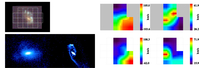}\\
 J033234.12-273953.5  (INC 1:1 $r_{\rm peri}$=0.4) & J033239.04-274132.4  (POL 1:1 $r_{\rm peri}$=0.4) \\
\includegraphics[height=2.9cm]{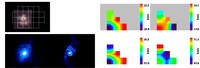}& \includegraphics[height=2.9cm]{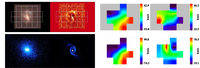}\\
 J033239.72-275154.7  (DIR 3:1 $r_{\rm peri}$=0.2) {\it Spin inv. for the small galaxy} &J033240.04-274418.6  (RET 3:1 $r_{\rm peri}$=0.2) \\
\includegraphics[height=2.9cm]{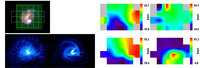}& \includegraphics[height=2.9cm]{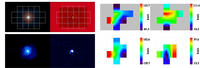}\\
 J033241.88-274853.9  (INC 1:1 $r_{\rm peri}$=0.4) &J033244.20-274733.5  (INC 1:1 $r_{\rm peri}$=0.2) \\
\includegraphics[height=2.9cm]{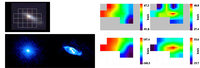}& \includegraphics[height=2.9cm]{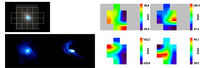}\\
J033245.11-274724.0  (INC 3:1 $r_{\rm peri}$=0.4) & J033248.28-275028.9  (INC 3:1 $r_{\rm peri}$=0.2) \\
\includegraphics[height=2.9cm]{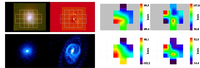}& \includegraphics[height=2.9cm]{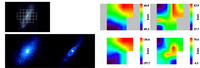}\\
 J033249.53-274630.0  (DIR 1:1 $r_{\rm peri}$=0.2) {\it Spin inv. for the main galaxy}& J033250.53-274800.7  (POL 3:1 $r_{\rm peri}$=0.2) \\
\includegraphics[height=2.9cm]{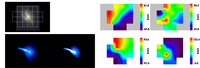}&\includegraphics[height=2.9cm]{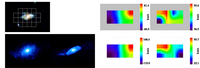}\\
\end{tabular}
\caption{continued}
\end{figure*} 

\subsection{Distribution of mass ratio, merger temporal phases and orbits}

In summary, we find that among 33 distant galaxies, 17 are robustly
and 9 are possibly reproduced by models of major mergers. In our 
simplified model, our goal is just to identify which  configuration
(phase, orbit, mass ratio,  \& pericenter, see Table 3) is able to
reproduce both morphology and kinematics. There are several 
possible biases in such an exercise and several of them have been 
discussed above. It is however interesting to examine the overall 
distribution of the configurations parameters that could reproduce the 
distant starbursts as mergers.

First, we may consider the mass ratio between the two interlopers. For each configuration in which the two interlopers can be identified, we have used z-band photometry to calculate the mass ratio (see Table 2). However many starbursts have been identified with merger remnants for which we derive the mass ratio from the modelling. Figure 3 (top) shows a distribution with two peaks at 1:1 and 3:1 (log($M_{2}$/$M_{1}$)=0 and -0.48, respectively), which are obviously  related to the adopted methodology.  Photometric estimates of the mass ratio can be done for mergers before the second passage (see Figures 1 and 2) when the two components can be separated. Nearly half of the sample possess photometric estimates of the mass ratio, and they draw a smoother distribution, ranging mostly from 0.25 to 0.65 for $M_{2}$/$M_{1}$ (Figure 3, middle). One may wonder how can be derived in such a way the mass properties of a disturbed dark matter component, especially for the minor interloper that is likely harassed during the event. \cite{Stewart09a} studied such configurations (see their Figure 2, right panel for gas-rich z=1 galaxies), and found that the $M_{2}$/$M_{1}$ stellar value ranges from 1/3 to 3/2 times the values for the dark matter, assuming stellar masses in the range of $10^{10}$ to $10^{11}$ $M_{\odot}$, respectively.  Figure 3 (bottom) shows the distribution of dark matter ratio after applying the correction suggested by \cite{Stewart09a}. The main difference between the top and middle/bottom panels of Figure 3 is the vanishing of the 1:1 peak: it is not surprising that equal mass mergers are rarer than 2:1 or 3:1 mergers, and indeed one can notice that a large fraction of 1:1 mergers are not robustly modelled. Both distributions are overwhelmingly dominated by major mergers \citep[all but the satellite infall][]{Puech07}. The overall distribution  shows the scarceness of events involving a galaxy more massive than the observed one, since those are rarer due to the exponential drop of the mass function towards the massive end. The quasi absence of minor merger may have a different meaning because minor encounters should be numerous at $z_{median}$=0.65 \citep[e.g.][]{Davies2009}. In fact minor mergers are expected to affect less and in a more sporadic way, kinematics, morphology and star formation \citep[see also ][and discussion in section 4.1]{Hopkins08}.  Overall the distribution of mass ratio seems consistent with a modelling of most distant starbursts as major mergers as shown in section 2.2.

\begin{figure}
   \centering
\includegraphics[width=9cm]{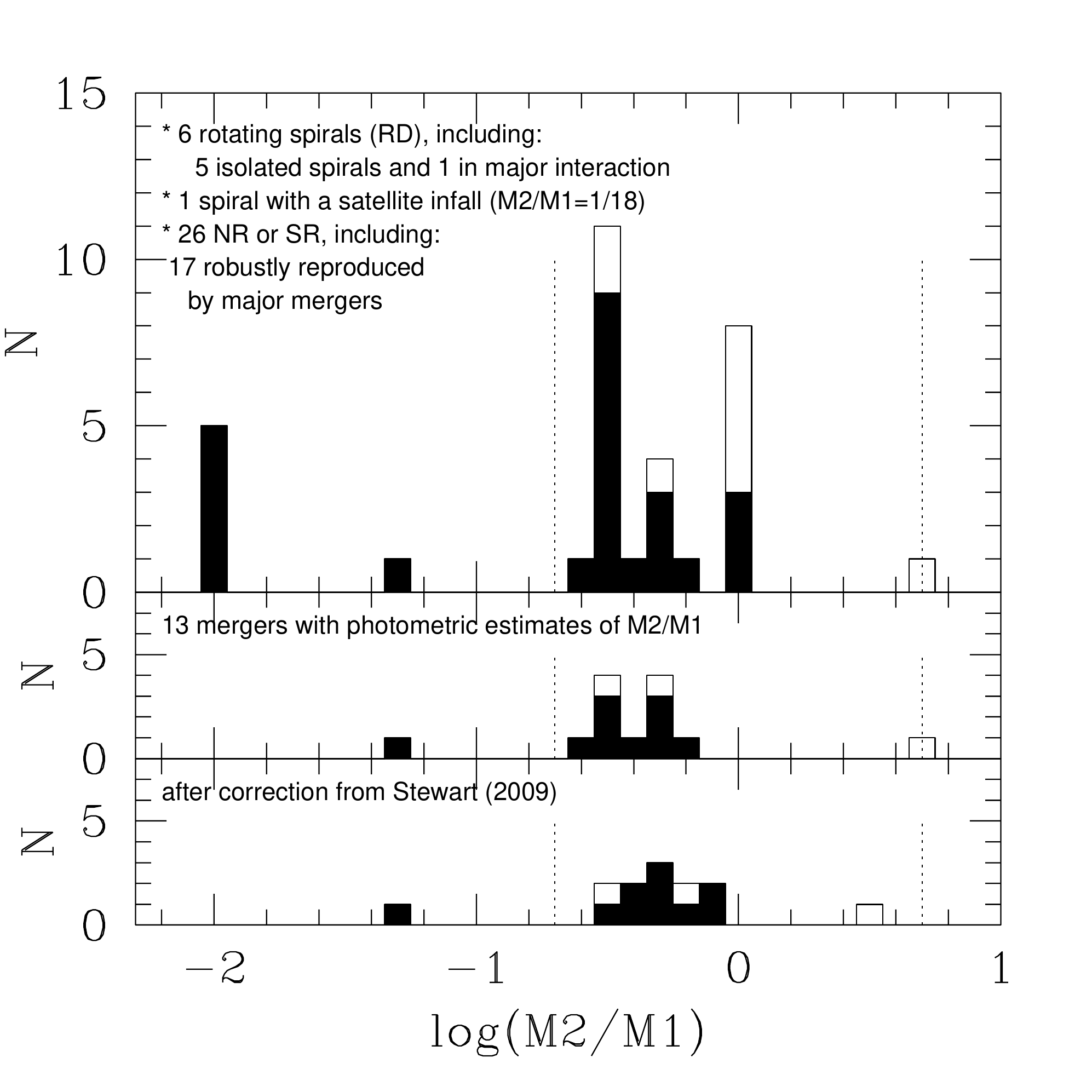}
   \caption{{\it Top}: distribution of the logarithm of the mass
     ratio, M1 representing the mass of the galaxy observed by
     IMAGES. The full black histogram corresponds to the 24 galaxies
     the nature of which we have robustly identified, including the 17 robust mergers, the
     5 isolated rotating disks (arbitrarily set at M2/M1=0.01), the spiral galaxy in interaction and the galaxy with a satellite infall. The
     vertical dotted lines represent the limit of major mergers (mass
     ratio between 5 and 1/5). {\it Middle}: same as above but for
     galaxies with separated components (see Fig. 2) for which we have
     been able to estimate the mass ratio using the z-band photometry
     from ACS, which corresponds to the rest-frame V-band. {\it Bottom}: same as above but after applying the correction by \cite{Stewart09a} to recover the dark matter M2/M1 ratio. We have generalised this correction factor to values of  M2/M1 (stellar) different than 0.3.} 

              \label{Fig3}
    \end{figure}
 
 Figure 4 shows how the modelled
galaxies are distributed during the various temporal phases of the
merger.  The combination of constraints from large-scale kinematics and from detailed
morphology generally leaves few doubts about the merger-phase. For example for J033224.60-274428.1 (see Fig. 2) the collision
could not be reproduced by a second passage, that would not fit both
the morphology and the dispersion peak location in direct or inclined orbits.
Furthermore, we believe that most galaxies have their phases quite
robustly identified. This is even true for several of the nine "possible" mergers, including J033220.48-275143.9 for which the disturbed morphologies of both components evidences a phase between the first and the second passage. Similarly, if J033214.97-275005.5, J033225.26-274524.0, J033228.48-274826.6, J033234.12-273953.5, and J033244.20-274733.5 are really mergers their highly distorted morphologies or their compactness is difficult to understand if they were not near the fusion stage. For J033238.60-274631.4 (see Fig. 1, right) it is very plausible that the interaction is before a first passage, although it is unclear whether it could be a simple fly-by or a first stage of a merger.

The result shows a relatively equal distribution
of the merger-phases in the IMAGES sample of distant-starburst
galaxies. In Fig. 4, we have added the 5 rotating spirals with warm
gaseous-disks from their low values of V/$\sigma$ \citep[see][]{Puech07}. These galaxies could well correspond to a very last phase: a
relaxation after their disks has been rebuilt. This would also explain why
these galaxies are forming stars efficiently, mostly in their
outskirts \citep[see][]{Neichel08}: they are still fed by the late
infall of the gas particles that have been expelled at larger radii by
the collision.

Figure 4 draws an evolutionary sequence in which {\it all} distant
starbursts can be identified to a major merger phase and are
subsequently modelled. This sequence is complementary to that drawn
by \citet{Hammer05} (see their Figure 6). Notice that distant starbursts represent a significant fraction of distant galaxies as they correspond to 60\% of the
galaxy population at $z_{median}$=0.65 \citep{Hammer97}.

\begin{figure}
   \centering
\includegraphics[width=9cm]{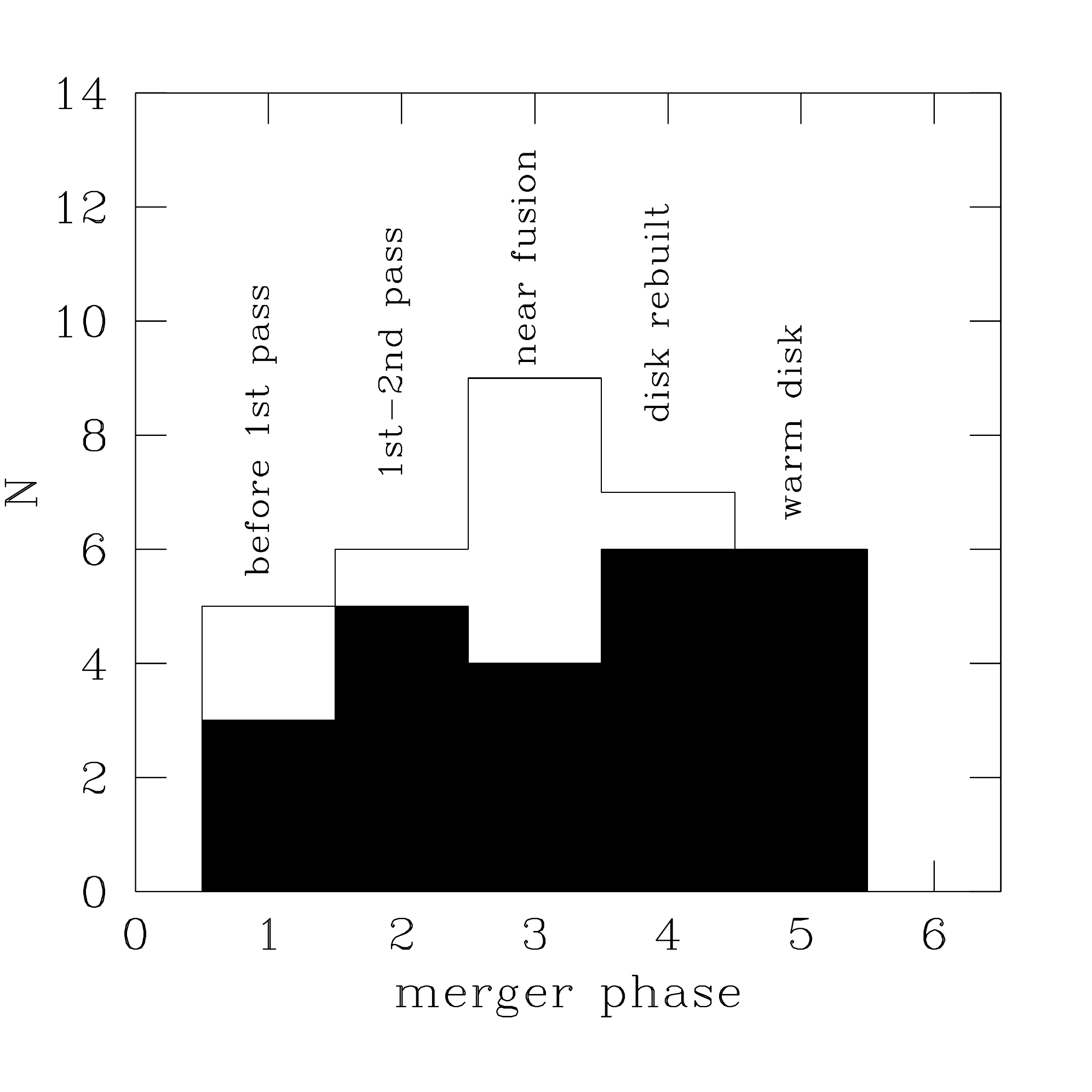}
   \caption{Distribution of the merger-phases for the modelled distant
     starbursts. Full black histogram corresponds to the same galaxies than in Fig. 3 (robust mergers/interactions and isolated rotating spirals). Phase 1
     corresponds to the approaching phase until the first passage, phase 2  corresponds to the
     time elapsed between the first and the second passage, and phase 3 to the time 
     after the second passage and
     before the elaboration of the rebuilt disk. During phase 3 the
     galaxy may take the appearance of a chaotic morphology  or a central starburst , always
     accompanied with chaotic velocity field and dispersion peaks
     clearly offset from the mass center. Phase 4 correspond to the
     disk rebuilding phase, for which the disk is detected, the
     rotation is evidenced but could be offset from the main optical
     axis, and the dispersion peak(s) is (are) closer to the mass
     centre. The last phase corresponds to rotating
     spirals with regular velocity fields and a dispersion peak centred
     to the mass centre \citep[e.g][]{Flores06}.} 

              \label{Fig4}
    \end{figure}
    
Conversely to the mass ratio and the merger phase, an accurate determination of the orbit is much more difficult, which is possibly 
due to the adopted methodology. Indeed we may have lost some 
configurations especially for phases during or after the fusion. 
Another difficulty is a possible degeneracy between different orbits. 
For example, the galaxy morphology of J033234.12-273953.5 may be well 
reproduced by a retrograde merger (even better than with the adopted 
inclined orbit) although we have not been able to reproduce the location 
of the dispersion peak. Fig. 5 shows the distribution of orbits within 
the sample of 26 mergers or possible mergers plus the rotating spiral in interaction for which the orbit is not constrained at all (labelled N/A).  Only 12 galaxies have 
their orbits robustly determined, i.e. galaxy morphologies and 
kinematics cannot be reproduced by other orbits. It also shows a lack 
of direct and retrograde orbits, which could be also related to the 
methodology. Indeed, we have tried to reproduce both kinematics and 
morphological details such as starbursting bars, rings and helicoidal 
structures that are generally associated to inclined and polar orbits. 

\begin{figure}
   \centering
\includegraphics[width=9cm]{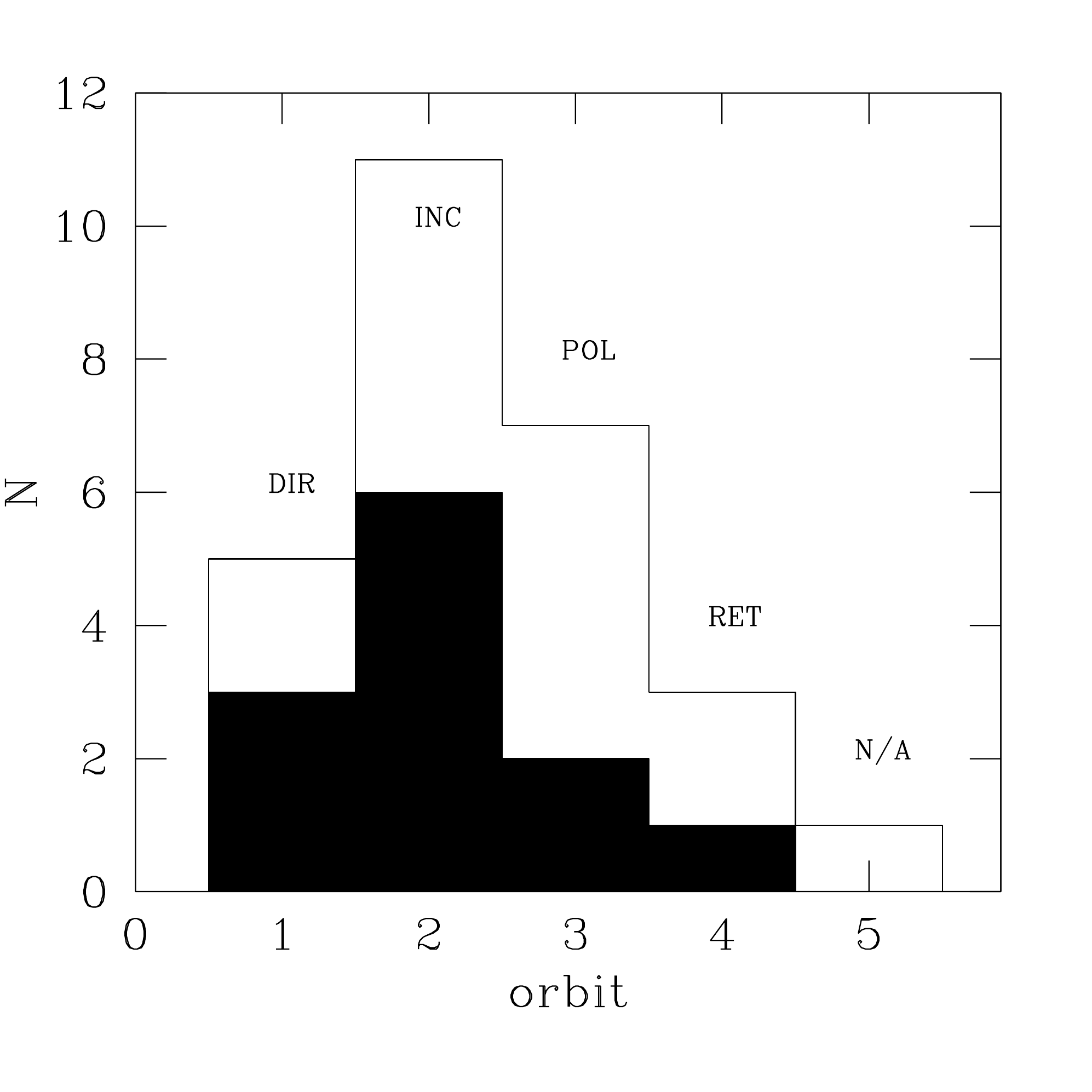}
   \caption{ Distribution of the interaction orbits adopted to model
     the 26 mergers plus one interaction in the sample of 33 distant starbursts observed
     with IMAGES. Full black histogram corresponds to the 12 objects
     for which we have robustly identified the orbit. The x-axis represents the different orbits displayed in the Figure.}

              \label{Fig5}
    \end{figure}

Only detailed modelling may solve the degeneracy in the orbit determination and help to verify
whether the orbital distribution is consistent with the hierarchical model
of galaxy formation. In this section we have just demonstrated that
for the many distant starbursts having complex morphologies and
kinematics, most of them can be reproduced by a simple modelling of
major mergers. The next step is to estimate whether this is just a
coincidence, or if mergers may explain the elaboration of spirals, as
suggested in Fig. 4. \citet{Robertson06} and many subsequent
studies \citep{Hopkins09a,Governato09} have shown that
within the conditions of sufficient gas richness -generally assumed to
be over 40-50\%- major mergers lead to the rebuilding of a significant disk.

\section{Gaseous content of distant starbursts}

\subsection{Gas and stellar masses}
The gas content of distant galaxies is still poorly known. Large interferometric sub-mm baseline (ALMA) will be essential especially in providing estimates of the molecular gas that is related quite intimately with star formation \citep{Gao04}.  We may however make use of galaxies in the local universe, for which the surface densities of star formation and gas are observed to follow a Schmidt-Kennicutt law,
 $\Sigma_{SFR}$ $\sim$  $\Sigma_{gas}^{1.4}$, over more than 6 orders of magnitude in  SFR
\citep{Kennicutt98}. This empirical relation is usually explained by a model in which the SFR scales with density-dependent gravitational instabilities in the gas. 

 It is plausible that $z_{median}$=0.65 starbursts, including those 
involved in mergers are following the same relation as local
starbursts and mergers. Gaseous masses have been estimated by \citet{Puech09b} by assuming that distant starbursts followed the 
Schmidt-Kennicutt relation. It generates gas surface densities that 
range from 10 to 250 $M_{\odot}$ $pc^{-2}$, i.e. intermediate between
normal spirals and ULIRGs.  Such a method has been already used to
derive the gaseous mass fraction of very distant galaxies \citep{Erb06}.  Because the local estimate of the gas surface density is 
independent of the IMF, \citet{Erb06} noticed that by using the same 
Salpeter IMF adopted by  \citet{Kennicutt98}, the derived gaseous mass 
is in principle an IMF independent quantity.  By accounting for the 
uncertainties of the SFR derived from Spitzer/MIPS 24$\mu$m
measurements \citep[Salpeter IMF][]{Kennicutt98}, \citet{Puech09b}  
derived error bars ranging from 0.04 to 0.4 dex for gaseous mass estimates.

 Ironically, the wealth of UV to mid-IR data in IMAGES does not
 provide yet a better accuracy in estimating the stellar
 masses. Indeed at all wavelengths most of the emission are due to 
 massive stars and not to the main sequence stars that constitute the 
 bulk of the stellar mass. The results are heavily dependent on the
 IMF, the assumed history of star formation and on various recipes 
 for the extinction law. \citet{Puech08} (see their Appendix A) 
 have analysed the systematic effects in adopting different schemes 
 for the stellar mass estimates. They have adopted stellar masses 
 ($M_{stellar, B03}$) calculated from rest-frame K-band magnitudes, 
 using the methodology of \citet{Bell03} to correct for massive 
 stars and assuming a "diet" Salpeter IMF. By comparing the evolution 
 of $M_{stellar}$/$L_{K}$ to that from comparable studies, they found 
 that at $z_{median}$=0.65, the \citet{Bell03} method
 overestimates the stellar mass by $\sim$ 0.1 dex when compared to 
 that derived from \citet{BC03}(BC03). Furthermore \citet{Maraston05} showed that by including TP-AGB stars, the 
 $M_{stellar}$/$L_{K}$ is overestimated by an additional $\sim$ 
 0.14 dex. Thus stellar-mass estimates from \citet{Bell03}
 ($M_{stellar, B03}$) appear to be {\it maximal estimates for the 
 stellar masses} of the IMAGES galaxies considered here. It also 
 gives an approximation of what could be estimated from a combination 
of the Salpeter IMF and \citet{BC03} models (see arrows 
in Fig. 6): the IMF effect (+0.15 dex) is compensated by the
 overestimate due to the \citet{Bell03} methodology when compared 
to \citet{BC03}. 
  Note that the Salpeter IMF is the maximal IMF and also violates the 
maximum disk constraints for local spirals (see e.g. Bell et al.,
 2003). Adoption of other IMFs \citep[for example,][]{kroupa02} unavoidably 
confirms that  $M_{stellar, B03}$ overestimates the stellar mass
 \citep[e.g.][]{Bell03}. Note that the above applies for the 
whole population of galaxies considered here, and that the estimate of 
stellar masses for individual galaxies is still challenging.

 Figure 6 shows how the gaseous masses are distributed against the maximal estimates of the stellar mass (see also Table 3). All distant starbursts but one show gas fractions intermediate between the local values and 50\%, and there is a correlation between their  stellar and gaseous masses. Such a correlation may be expected as we have selected starburst galaxies on the basis of their [OII] equivalent widths and a proxy of their stellar masses (($W_0(OII)\ge$ 15\AA~ \& $M_{J}(AB)<$ -20.3, see IMAGES-I). 

\begin{figure}
   \centering
\includegraphics[width=9cm]{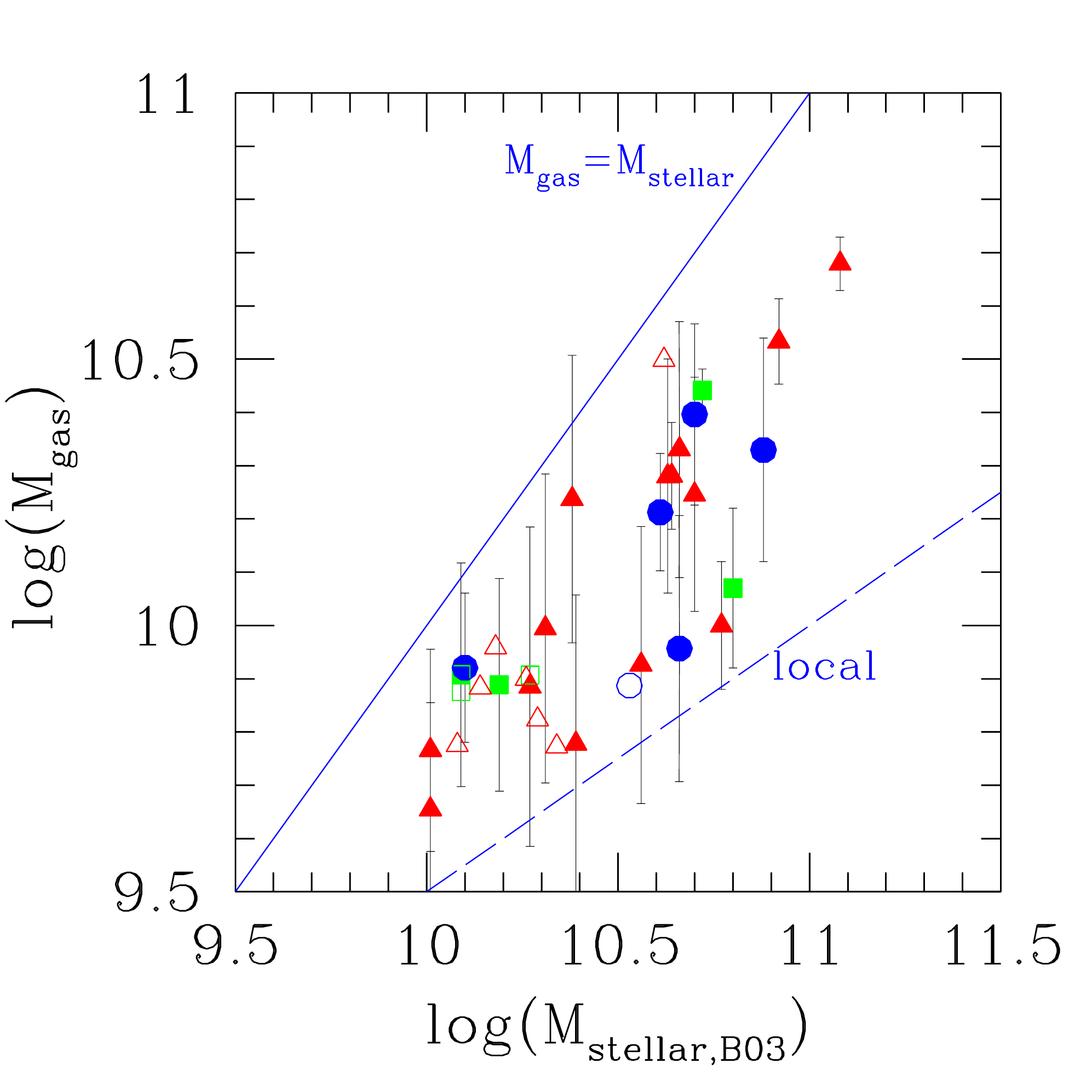}
   \caption{Gas mass derived from \citet{Puech09b}  as a function of the
     stellar mass obtained from the \citet{Bell03}
     prescriptions. The arrows in the upper left corner indicate how
     the stellar mass may be affected by different prescriptions of
     the galaxy synthesis population models (BC03  and M06 indicate
     the change by adopting Bruzual and Charlot, 2003 and Maraston,
     2006 models, respectively), or of the IMF (S55  and K02 show the
     change due to the adoption of an IMF from Salpeter, 1955 or from
     Kroupa, 2002, respectively). The full and dashed lines indicate
     a gas fraction of 50\% and a gas fraction from local galaxies  \citep[HI  gas,][]{Schiminovich08}, respectively. The various symbols correspond to
     the morpho-kinematic classes defined in section 1: rotating
     spirals (blue circles),  heavily non-relaxed systems (red
     triangles) and semi-relaxed systems (green
     squares). Only error bars lower than 0.3 dex have been
     represented in this plot, and values with larger error bars are
     represented as open symbols.} 

              \label{Fig6}
    \end{figure}

\subsection{Gas fraction in distant starbursts and their progenitors}

The median gas fraction of the sample is 31\%$\pm$1\%, which is much
larger than the corresponding fraction for the Milky Way (12\%, Flynn
et al. 2006) or for M31 \citep[5\%,][]{Carignan06}. It is
approximately twice the value found by ALFALFA \citep{Schiminovich08}. Different prescriptions for synthesis models or IMF would lead
to larger gas fractions in the distant sample: for example, combining
a Maraston model with a Kroupa IMF would divide the stellar mass by a
factor $\sim$ 2, and thus the median gas fraction would become 47\%.
 It is important to stress that we generate {\it
  minimal gas fractions} by applying maximal stellar masses. An independent confirmation of higher gas fractions in distant galaxies is provided by \cite{Rodrigues08} who studied the O/H abundances in the gaseous phases of similar distant starbursts. They found that gas fraction should reach $\sim$30\% at z$\sim$ 0.65 if one assumes $\sim$10\% the gas fraction in present-day galaxies. Given the large uncertainties in our estimates of gas fraction, an independent confirmation is certainly reassuring.

Figure 7 shows how the gas fractions are distributed in the different
merger phases defined in section 3. 
There is no specific trend between the two quantities.
In the frame of
a merger scenario, this is not unexpected because it results from the
balance between two effects. Let us consider two sets of z=0.6 galaxies, one (hereafter called A set) including galaxies in later merger stage (after the fusion) and the second (hereafter called B set) with galaxies before the first pass. By construction, progenitors of A galaxies are lying at higher redshift and should have lower luminosities (and masses) than B galaxies, and then should have, on average, a significantly higher gas fraction. This is somewhat compensated by the fact that A galaxies have had a supplementary star formation that is induced during the merger, which has transformed a part of the gas into stars, reducing the gas fraction. Given the small statistics (5 to 7 galaxies in each phase), we believe that Figure 7 is consistent with the merger hypothesis. 
Indeed the main question is whether the
progenitors of the observed galaxies are sufficiently gas-rich to lead
to disk rebuilding.

There are two possible means to estimate the gas fraction in merger progenitors. \citet{Hammer09b} have modelled the stellar populations from deep spectroscopy of a merger remnant assumed to be in a disk rebuilding phase. They found that $\sim$ 50\% of the observed stellar mass had been formed during the merger by comparing the stellar population ages to the merger dynamical time. This implies that the progenitors of this system were on average more than 50\% gas-rich, supporting the evidence that this system is rebuilding its disk after a major merger. Here we use the characteristic doubling time, $T_{SFR}$=$M_{stellar, B03}$/SFR, to estimate the stellar mass that has been formed during the event, and thus the gas fraction in their assumed progenitors. Assuming that our maximal stellar-masses are a good representation of the stellar mass for a Salpeter-IMF and a \citet{BC03} synthesis model, this quantity is IMF independent, as our SFR estimates have been done using the \citet{Kennicutt98} calibration (Salpeter-IMF) to the IR luminosities \citep[see details]{Puech09b}.

Figure 8 gives the distribution of the characteristic stellar mass doubling
times for the 34 IMAGES starbursts. It is remarkable that their median
value takes its minimum near the fusion, which is expected in every
model of mergers. We have used the models shown in section 3 to
estimate the time each galaxy spends in each of the phases (see time values in Table 3). It assumes a
rotating period time of $t_{rot}$=1.2 $\times$ 0.25 Gyr for a galaxy
with the mass of the Milky Way \citep[see][]{Barnes02}. We then scale the
merger time with the observed baryonic mass (assuming $M_{MW,
  baryonic}$=5.5 $10^{10}$ $M_{\odot}$) and also apply a correction
for the merger mass ratio as described in \citet{Jiang08}. Besides
this, we calculate for each phase the median $T_{SFR}$ that is
considered to be its effective star formation time.  For a given
starburst assumed to be in a given phase, we may calculate the
fraction of gas that has been transformed into stars during each
previous phases of the merger. Table 3 (column 10) gives the time spent by each galaxy in the previous merger phases, that, after combination with the median $T_{SFR}$, provides us with an estimate of the gas mass that has been transformed into stars during the merger.


Figure 7 also shows the resulting (median) distribution of the progenitor
gas-fractions (see also Table 3). In 75\% of the progenitors, the gas-fraction is in
excess of 40\% or of 58\% if we adopt a Maraston-Kroupa
combination. The high gas fraction in progenitors is robust
because:\newline 
- it only requires a modest star formation in the merger to reach large values since we already find a high fraction of gas (median 31\%) in the observed starbursts ; \newline
-  the gas-to-star transformed mass during the merger is independent of the observed mass and thus of the IMF choice, since both $T_{SFR}$ and the merger time scale with mass.\newline


\begin{figure}
   \centering
\includegraphics[width=9cm]{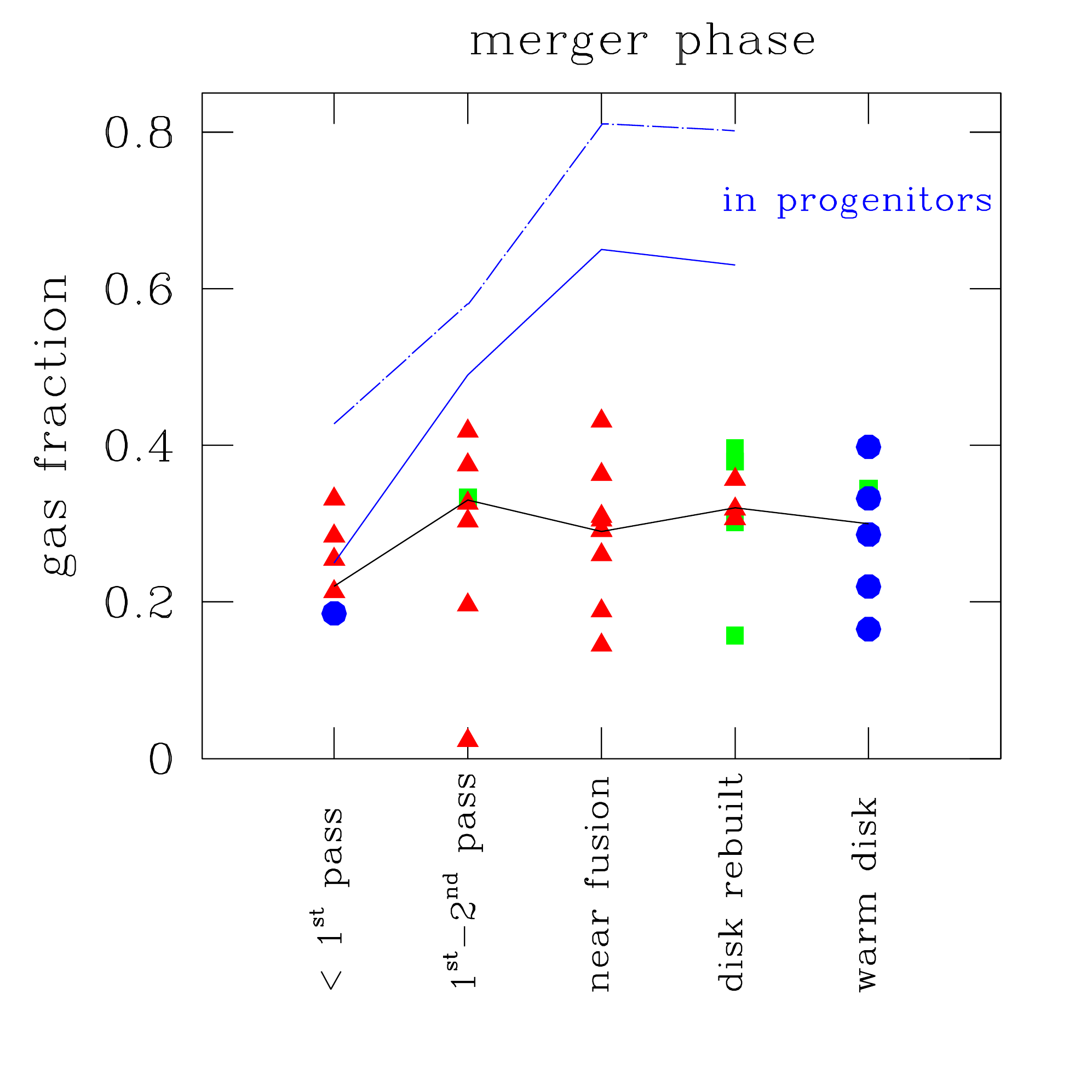}
   \caption{Gas fraction distribution as a function of the merger
     phases. The former corresponds to a minimal value of the gas
     fraction in the sample of IMAGES starbursts (see text). The various 
     symbols correspond to the  morpho-kinematic classes, as
     in Fig. 6. Note the
     presence of the massive dry merger at phase=2 (see the inserted image in Fig. 6) with a
     very low gas fraction \citep{Yang09}. The full blue line shows the median
     gas-fraction in the progenitors derived from the gas consumption during the merger (see text). The dot-long-dashed line shows
     the same, but for a combination of \citet{Maraston05} models and
     Kroupa IMF. The black solid line gives the median of the gas
     fraction values. }

              \label{Fig7}
    \end{figure}

\begin{figure}
   \centering
\includegraphics[width=9cm]{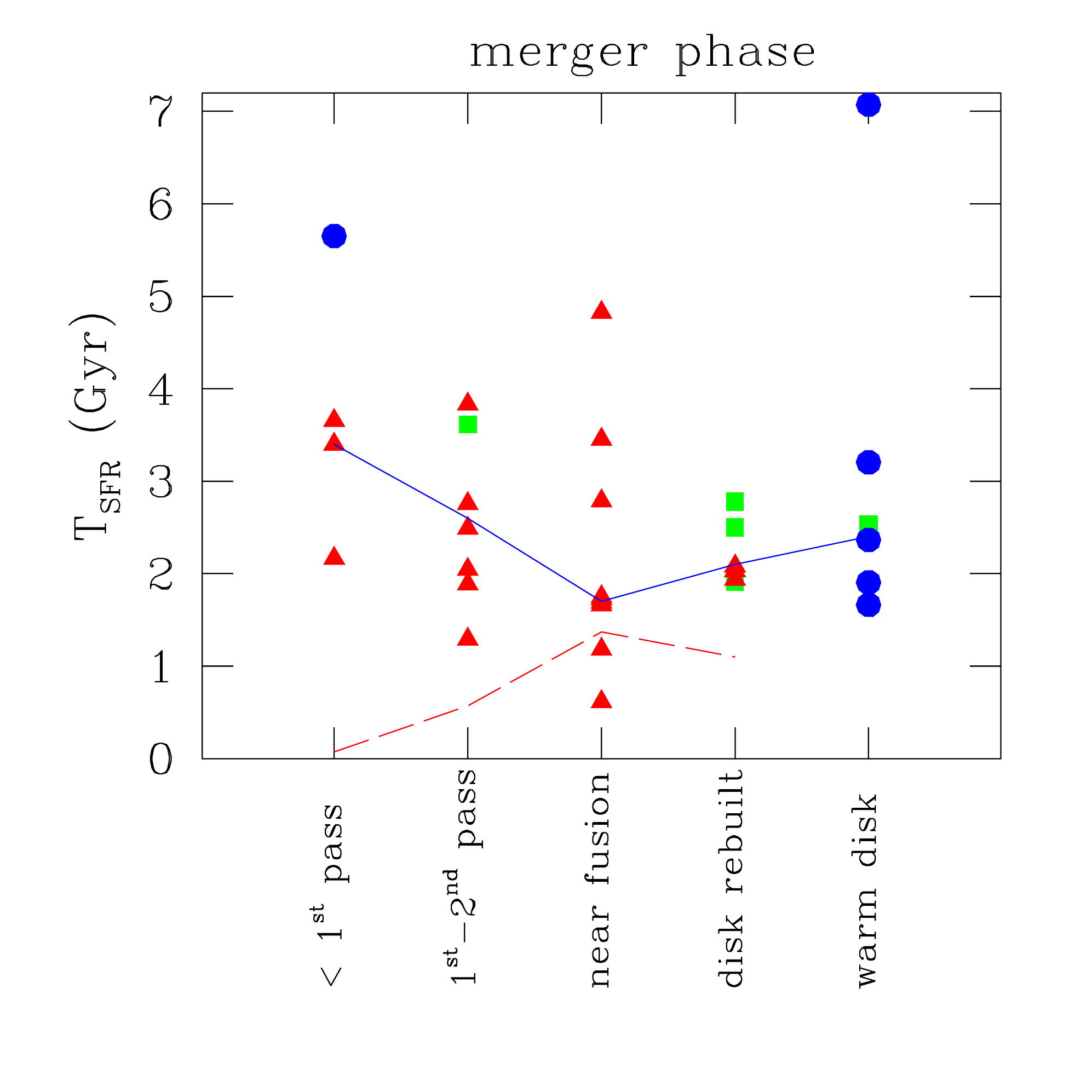}
   \caption{Characteristic stellar mass doubling times as a function
     of the merger phases, as they are described in Fig. 4. The various 
     symbols  corresponds to morpho-kinematics classes as in Fig. 6. The
     full blue line indicates their median value and the dashed red
     line indicate the time during which the observed starbursts are
     involved in the merger (median value for each phases 1 to 4), as
     obtained from the simulations. It slightly decreases at phase 4
     because the masses of the 4 rebuilding disks (median value, 2.1
     $10^{10}$ $M_{\odot}$) are smaller than the mass of the 9
     galaxies in phase 3 (median value, 6.3 $10^{10}$ $M_{\odot}$). } 

              \label{Fig8}
    \end{figure}

  \section{Discussion}
  \subsection{A self-consistent explanation of the galaxy transformation during the last 6 Gyrs }

 In section 2 we argued that distant starbursts have morphologies and kinematics consistent with
 major mergers or their remnants. In section 3 we show that they
 have large gas fractions, and that their progenitors would have to have gas
 fractions above 40-50\% to account for the stellar
 mass produced during the merger. Therefore most starbursts at
 $z_{median}$=0.65  -- those with anomalous morphologies and
 kinematics --
 are consistent with gas-rich merger phases leading to rebuilt
 disks.

 Our interpretation of the morpho-kinematic evolution (see
 Table 1) is then straightforward: $\sim$ 6 Gyrs ago, 46\% of the
 galaxy population was involved in major mergers and most of them
 (75\% $\times$ 46\%= 35\%)
 were sufficiently gas rich to rebuild a 
 disk. Those can be considered as progenitors of the present-day numerous spirals -- although this deserves a careful analysis of the
 exchanges of angular momenta -- while the others could be progenitors
 of E/S0 and of the scarce population of massive irregulars at the
 present-epoch \citep[$\sim$ 10\%, see ][]{Delgado09}. Thus as much as half of the present-day spirals 
 are coming from disk rebuilding from recent
 mergers, the other half being already assembled into quiescent or
 warm disks at  $z_{median}$=0.65 (Table 1).

 More statistics are certainly needed to obtain a more precise
 estimate of the amount of gas that has been consumed during the
 different merger phases. The median time spent in each merger phase
 ranges from 0.5 to 1.4 Gyr (see Fig. 8 and also Table 3): the scenario naturally
 explains why distant starbursts show a so important contribution of
 intermediate-age stars revealed by their very large Balmer absorption
 lines in their spectra (e.g. \citealt{Hammer97,Marcillac06}, see also \citealt{Poggianti99} for another perspective in
 galaxy-cluster environments).  The median baryonic mass of the sample
 is 0.75 times that of the Milky Way. Their progenitors should be
 galaxies at larger redshifts, approximately 1 Gyr earlier, i.e. at
 z$\sim$ 0.83. At such redshifts the large gas fractions in
 progenitors is not exceptional. Accounting for the gas consumed
 during the merger, the median stellar mass and gas fraction of their
 progenitors are 7.5 $10^{9}$ $M_{\odot}$ and 50\%,
 respectively. In present-day galaxies within this mass range, the gas
 fraction averages to $\sim$26\% for local galaxies \citep[from][]{Schiminovich08}, and it could be not exceptional that 7 Gyrs ago such
 galaxies had twice their present gas content.

 Improvements are also required to estimate the stellar masses since a
 proxy (absolute J-band magnitude) of the stellar mass has been used
 in this study to select our sample. Combination of realistic stellar
 population with different ages and metal content has to be performed
 on both the whole spectral energy distribution (from UV to near-IR)
 and the spectroscopic absorption lines (Lick indices). Nevertheless,
 we do find that all distant starbursts are consistent with major
 merger phases, and these sources are strong emitters in near-IR. It
 is unclear whether we may have missed a significant population of
 massive starbursts without strong emission in near-IR. Besides this,
 technical limitations (see IMAGES-I) have prevented us from measuring
 the kinematics of 3 starbursts, because their emission line region
 are too compact (see Fig. 9). Their optical morphologies are
 also consistent with mergers (see Fig. 9 and its caption).

 \begin{figure}
   \centering
\includegraphics[width=9cm]{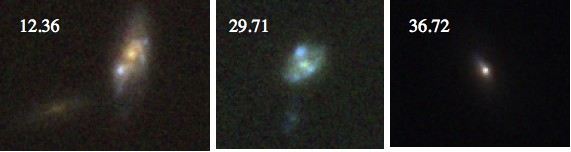}
   \caption{b+v, i and z combined images of three additional starbursts whose kinematics has not been detected due to spatial resolution. The galaxy in the middle panel show a ring and possibly the nucleus of the secondary interloper. Kinematics are needed to confirm it as well as to verify the nature of the two other starbursts that could be in close interaction and just at the nuclei fusion, respectively. }

              \label{Fig9}
    \end{figure}

 Figure 4 -- see also Figures 1 and 2 -- suggests that all distant
 starbursts are part of the duty cycle of the disk rebuilding scenario
 \citep[see][their Figure 6]{Hammer05}. It also includes the
 distant rotating spirals with high star formation and warm disks,
 which are the natural last phase of such a gas-rich merger event. During the elapsed
 time to z=0, they may transform their gas-into-stars and simply relax
 to form the present-day thin disks. During such phase, models predict
 that almost all the gas has reached the disk, which is confirmed
 by the fact that they lie on the same baryonic Tully Fisher relation as local
 disks \citep{Puech09b} . In passing, this brings a simple
 explanation of why some regular disks are LIRGs \citep{Melbourne06}.  
 
The spiral disk rebuilding scenario explains the changes
observed in the galaxy population, the density evolution of
star formation and stellar mass, the dispersion of the
evolved Tully Fisher relation \citep{Flores06,Puech09b} as well as the strong evolution of the metal abundances of
their gas phases \citep{Rodrigues08}. It is also consistent with
the observations of galaxy pairs at $z\sim$ 0.6. The most robust
estimate is that 5$\pm$1\% of galaxies are in pairs at that epoch \citep[see e.g.][ for a review]{Bell06}: in our sample we identify only 4
galaxies which could have been identified as well separated pairs. It
corresponds to a pair fraction of 4/33$\times$0.6= 7$\pm$3\% 
(see Table 1 in which 0.6 is the fraction of starburst galaxies) in the $z_{median}$=0.65
population. Note that among these pairs, three are in phase 2,
i.e. between the first and the second passage. Last but not least,
major mergers may explain why the ionised gas radius is larger than
the optical radii, especially for the starbursts with small optical
radius  \citep{Puech09b}. Indeed, during a close encountering and nuclei fusion phases, the
gas is heavily shocked by the collision and could be ionised this way
 \citep{Puech09a}. Another alternative is that dust-enshrouded clumps may ionise the gas while 
they cannot be detected at visible wavelengths; this phenomenon has been identified in a compact dust-enshrouded disk \citep[see][]{Hammer09b}.

Cosmological simulations confirm the importance of the gas in mergers.  By accounting for the gas, \citep{Stewart09} found that almost all galaxies may have experienced a merger since z=2.  A comparison  to IMAGES observations would be highly desirable to sharpen this prediction.  Those may be consistent with the whole formation of the Hubble sequence: a formation of spirals by the numerous gas-rich mergers and of massive ellipticals by gas-poor mergers. It is now time to study the formation (or re-formation) of the present-day spiral galaxies and of their substructures and to catch a glimpse on how the whole Hubble sequence may have been formed.

\subsection{How other mechanisms contribute to the galaxy transformation?}
Numerous studies have attempted to describe the above evolution by
assuming different mechanisms. It has been argued that only a small fraction of the star formation is triggered by major interactions \citep{Robaina09,Jogee09}. Although these studies are based on a large number of objects, they do not possess the kinematical information that is crucial to evaluate the presence or absence of merging. Indeed \cite{Neichel08} found that most massive galaxies showing irregularities or compactness have also anomalous velocity fields. Other mechanisms than merging have then to explain not only morphological irregularities but also their non-relaxed kinematics.

We have examined the spectra \citep[see e.g.][]{Rodrigues08} of 20 IMAGES starbursts to investigate
whether outflows can be detected. An outflow may lead to significant
differences between the velocity of the emission-line system and that
of the absorption-line system \citep{Heckman00}. We do find
differences in only 3 (J033214.97-275005.5, J033224.60-274428.1 and J033225.26-274524.0) within 19 objects, at the level of $\sim$ 100
km/s.  Thus stellar feedback mechanisms are unlikely to considerably affect distant starbursts, which is understandable as all of them are relatively
massive with baryonic masses in excess of $10^{10}$ $M_{\odot}$. The
absence of minor mergers in our sample (only one case with 18:1
mass ratio) could be simply explained by their considerably lower
efficiency. Due to their lower impact and their longer duration \citep{Jiang08}, they are considerably less efficient to activate a
starburst, and to distort morphologies and kinematics or then they do
it in a somewhat sporadic way \citep[see][]{Hopkins08}. To explain
both the stellar mass assembly and the peculiar morpho-kinematics would
probably need an extremely large (non-observed) rate of minor mergers,
that is certainly not consistent with the large angular momenta of
present-day disks (see e.g. Maller \& Dekel, 2002).

 \section{Conclusion: a scenario to explain the elaboration of the Hubble sequence?}

 In this paper we considered the possibility that the formation of the
Hubble sequence
relies to a large extent on past merger events. We used a sample of
objects around z=0.65 for which we have both morphology from the HST
and high quality kinematics (both velocity field and 2D velocity
dispersion maps) and compared them with simulation results varying the
viewing angles to obtain the best fits.
Although we can of course not prove that the origin of these objects is
indeed a merger, we can safely say that their observed properties are
well compatible with them being merger or their remnants. We want to stress
that this result was reached using all available data,
i.e. our comparisons included morphology, mean velocities and
dispersions. This strengthens our proposal very considerably.
 
A merger origin of the Hubble sequence is very intimately linked to
the disk rebuilding scenario, which has been succesfully argued both
from the observational point of view and from simulations \citep{Barnes02, Springel05b, Hopkins09a}. Our quantitative estimates of gas
fractions argue that more than a third of the galaxy population about 6 Gyrs
ago was sufficienty gas rich to rebuild a disk after the merger.
Thus our work argues that a merger origin of the Hubble sequence,
although it has not yet been proven, is a plausible alternative or channel for the formation of large disks in grand-design spirals. 
 
Half of the present-day spirals being in merger phases at $z_{median}$=0.65 naturally implies that most and probably all were shaped during gaseous-rich mergers at earlier epochs. These mergers at $z_{median}$=0.65 have generally begun 1 Gyr earlier, i.e. at z= 0.835. We may expect, that a similar amount of mergers had occurred a further 1 Gyr ago (from z=0.835 to z=1.07): then almost all spirals may have been rebuilt from their last major merger, during the last 8 Gyrs. This may apply to M31 \citep{Hammer07}, but the Milky Way appears quite exceptional as its properties imply a last merger only at much earlier epochs (10 to 11 Gyrs ago). The Milky Way quiescent history is well illustrated by its exceptional pristine halo; combined with its lack of angular momentum and stellar mass, this may simply indicate that our Galaxy has exceptionally avoided any major merger during a large fraction of the Hubble time  \citep{Hammer07}.

A considerable task is thus awaiting us. We have to relate the distant 
starbursts to local galaxies by modelling in detail all the distant galaxies for which we possess detailed morphologies and kinematics, i.e. about 100 galaxies. Although large, such a number is barely sufficient to describe the large variety of merger configurations. With such a modelling we will be able to derive the end-up properties of each starburst, by modelling their evolution 6 Gyrs later, and verify whether they are consistent with the present-day distribution of galaxies within the Hubble sequence.

In order to make better comparisons between observations and simulations
it would be useful to have a larger library of simulations covering longer evolution times and larger gas
fractions. A better coverage of the parameter space would also be very
useful, although from our experience, it will take some time before being able to recover detailed modelling of each of the observed galaxies. Such a library (e.g. 'GALMER', see \citealt{DiMatteo2008}) may be useful as it could be adapted to the IMAGES observational parameters.

Nevertheless, it was possible to reach a number of
conclusions. It is instructive to note how frequent structures
such as bars and rings are. We observe 9 bars and 6 rings in
our sample of 34 galaxies, and they have colours consistent with young
or intermediate-age stars, and as such they could have been formed during the merger. Figure 10 evidences that such structures are still persistent at late merger phases.
The fraction of barred galaxies that we find is compatible with that found by \cite{Sheth08} for the same redshift range. Bars and rings are also observed in
local galaxies and may well be triggered from the earlier interactions and
merging. For example, it is known that interactions can trigger bar
formation \citep{noguchi87,gerin90,Steinmetz99}. Also polar encounters can create rings,
pseudo-rings, or spirals, which have characteristics similar to the
observed ones \citep[see also][]{berentzen03}. Finally spiral patterns can
also be generated by encounters, as has been shown by both simulations
and observations \citep[e.g. ][and references therein]{Toomre81,gol78,gol79,DonnerES91,Tutukov06}.    
Adjustment of the models is clearly a central issue. The helicoidal
structure found in many distant starbursts (see examples in Figure 2) is likely due to the central torque described by \citet{Hopkins09a}, and
it may regulate the angular momentum transfer. This structure is preponderant in the nuclei fusion phase
and seems present in 
galaxies in later phases \citep{Hammer09b}. Its efficiency in regulating the
bulge-to-disk ratio is likely considerable \citep[see][]{Hopkins09a}
although larger statistics are mandatory to verify this prediction and
its actual role.    

 \begin{figure}
   \centering
\includegraphics[width=8cm]{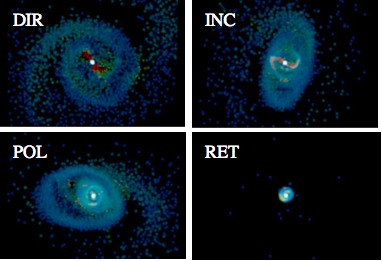}
   \caption{Gas distribution in merger remnants from \citet{Barnes02} for
     3:1 mergers, and for four different orbits, at the end of the
     simulations. }

              \label{Fig10}
    \end{figure}

\begin{sidewaystable*}

\caption{Table 3: general properties of the 33 distant starbursts, of their progenitors whether they can be assumed to be reproduced by a major merger, as well as merger parameters adopted after the modelling.}
\begin{tabular}{rrrrr rrrrr rrrr}\hline
     GOODS id   &        z  & morph$_{kin}$$^1$  &     Mgas  &     M$_{stellar}$ &  f$_{gas}$ & M2/M1$^2$  &  phase$^3$   &   orbit$^4$ &   dTmerger$^5$ &    Tmerger$^6$  &  Tsfr[Gyr]& f$_{gas}$ (prog)$^7$ & M$_{stellar}$ (prog)$^8$\\\hline
J033210.25-274819.5    &    0.6087    &       4 &  6.658e+09 &   1.95e+10 &     0.2545  &      0.38  &         1   &        1  &   0.09723  &    0.7134  &     3.398 &     0.2759 &   9.47e+09 \\
J033212.39-274353.6    &    0.4213    &       1 &  1.632e+10 &  4.074e+10 &      0.286  &         -  &         5   &        -  &         -  &         -  &     2.366 &          - &      -     \\
J033213.06-274204.8    &    0.4215    &       3 &  7.735e+09 &  1.549e+10 &     0.3331  &      0.26  &         2   &        1  &    0.1646  &    0.6334  &     3.614 &     0.3685 &  7.333e+09 \\ 
J033214.97-275005.5    &    0.6665    &       4 &  3.413e+10 &  8.318e+10 &     0.2909  &         1  &         3   &        2  &     2.399  &     4.532  &     1.752 &     0.9096 &  5.302e+09 \\
J033217.62-274257.4    &    0.6456    &       4 &  1.726e+10 &  2.399e+10 &     0.4184  &       0.7  &         2   &        2  &    0.3429  &     2.062  &      1.29 &     0.4828 &  1.067e+10 \\
J033219.32-274514.0    &    0.7241    &       4 &  5.992e+09 &  2.455e+10 &     0.1962  &      0.55  &         2   &        2  &     0.483  &     1.527  &     3.834 &     0.3214 &  1.036e+10 \\
J033219.61-274831.0    &    0.6699    &       4 &   7.68e+09 &  1.862e+10 &      0.292  &      0.33  &         3   &        3  &    0.6233  &     1.016  &     2.788 &     0.4525 &    7.2e+09 \\
J033219.68-275023.6    &    0.5595    &       1 &  2.136e+10 &  7.586e+10 &     0.2197  &         -  &         5   &        -  &         -  &         -  &     3.205 &          - &      -   \\
J033220.48-275143.9    &    0.6778    &       4 &  9.086e+09 &  1.514e+10 &     0.3751  &    0.2222  &         2   &        3  &     1.141  &     8.422  &     2.492 &     0.6051 &  4.783e+09 \\
J033224.60-274428.1    &    0.5368    &       4 &  5.961e+09 &  1.202e+10 &     0.3315  &      0.55  &         1   &        2  &   0.07111  &    0.6948  &     2.166 &     0.3455 &  5.886e+09 \\
J033225.26-274524.0    &    0.6647    &       4 &  8.436e+09 &  3.631e+10 &     0.1885  &      0.33  &         3   &        2  &    0.9788  &     1.729  &     4.826 &     0.4773 &  1.169e+10 \\
J033226.23-274222.8    &    0.6671    &       3 &  2.763e+10 &  5.248e+10 &     0.3449  &      0.05  &         5   &        -  &      1-  &    -  &     2.536 &      - & - \\
J033227.07-274404.7    &    0.7381    &       4 &  7.932e+09 &   1.82e+10 &     0.3036  &      0.33  &         2   &        2  &    0.5716  &      1.01  &     2.758 &      0.432 &  7.421e+09 \\
J033228.48-274826.6    &    0.6685    &       4 &  1.907e+10 &  4.266e+10 &      0.309  &         1  &         3   &        3  &     1.403  &     2.385  &     1.182 &     0.6615 &  1.045e+10 \\
J033230.43-275304.0    &    0.6453    &       4 &   1.91e+10 &  4.365e+10 &     0.3044  &      0.33  &         3   &        3  &     1.373  &     2.425  &     1.664 &     0.6516 &  1.093e+10 \\
J033230.57-274518.2    &    0.6798    &       4 &  4.778e+10 &  1.202e+11 &     0.2844  &      0.33  &         1   &        1  &    0.6125  &     4.582  &     2.044 &     0.4134 &  4.928e+10 \\
J033230.78-275455.0    &    0.6857    &       1 &  9.051e+09 &  4.571e+10 &     0.1653  &         -  &         5   &        -  &         -  &         -  &     7.073 &          - &      -  \\
J033231.58-274121.6    &    0.7041    &       1 &  8.325e+09 &  1.259e+10 &     0.3981  &         -  &         5   &        -  &         -  &         -  &     1.904 &          - &      -  \\
J033232.96-274106.8    &    0.4681    &       4 &  4.517e+09 &  1.023e+10 &     0.3062  &      0.33  &         4   &        3  &    0.4033  &    0.5699  &     2.082 &     0.4182 &  4.291e+09 \\
J033233.90-274237.9    &     0.618    &       4 &  2.139e+10 &  4.571e+10 &     0.3188  &         1  &         4   &        4  &      3.06  &      3.66  &     1.939 &      0.801 &  1.061e+10 \\
J033234.04-275009.7    &    0.7016    &       3 &  7.519e+09 &   1.23e+10 &     0.3793  &         1  &         4   &        4  &    0.4955  &    0.5406  &     1.914 &     0.5024 &  4.932e+09 \\
J033234.12-273953.5    &    0.6273    &       4 &  1.763e+10 &  5.012e+10 &     0.2603  &         1  &         3   &        2  &     2.618  &     3.388  &     1.724 &     0.9644 &  1.206e+09 \\
J033237.54-274838.9    &    0.6637    &       1 &  2.492e+10 &  5.012e+10 &     0.3321  &         -  &         5   &        -  &         -  &         -  &     1.663 &          - &      -  \\
J033238.60-274631.4    &    0.6206    &       1 &   7.71e+09 &  3.388e+10 &     0.1854  &       0.5  &         1   &        5  &    0.0808  &     1.891  &     5.651 &     0.2047 &  1.654e+10 \\
J033239.04-274132.4    &    0.7318    &       4 &  7.641e+09 &   1.38e+10 &     0.3563  &         1  &         4   &        3  &     1.072  &     1.218  &     2.031 &     0.6324 &  3.942e+09 \\
J033239.72-275154.7    &    0.4151    &       4 &  9.879e+09 &  2.042e+10 &     0.3261  &      0.33  &         2   &        1  &    0.3866  &    0.8263  &     1.891 &     0.4101 &  8.936e+09 \\
J033240.04-274418.6    &     0.522    &       4 &  9.991e+09 &  5.888e+10 &     0.1451  &      0.33  &         3   &        4  &     1.381  &     1.878  &     3.453 &     0.5744 &  1.466e+10 \\
J033241.88-274853.9    &     0.667    &       3 &   8.07e+09 &  1.862e+10 &     0.3023  &         1  &         4   &        2  &     1.213  &     1.335  &     2.776 &     0.6409 &  4.792e+09 \\
J033244.20-274733.5    &     0.736    &       4 &  3.155e+10 &  4.169e+10 &     0.4308  &         1  &         3   &        2  &     1.498  &     2.829  &    0.6172 &     0.7408 &   9.49e+09 \\
J033245.11-274724.0    &    0.4346    &       3 &  1.175e+10 &   6.31e+10 &      0.157  &         1  &         4   &        2  &     2.381  &     2.892  &     2.781 &       0.96 &  1.496e+09 \\
J033248.28-275028.9    &    0.4446    &       3 &  8.076e+09 &   1.23e+10 &     0.3963  &      0.33  &         4   &        2  &    0.5944  &    0.7874  &     2.502 &     0.5398 &  4.689e+09 \\
J033249.53-274630.0    &    0.5221    &       4 &  5.923e+09 &  2.188e+10 &     0.2131  &      0.46  &         1   &        1  &   0.05323  &     1.327  &     3.657 &     0.2254 &  1.077e+10 \\
J033250.53-274800.7    &     0.736    &       4 &  5.832e+09 &  1.023e+10 &      0.363  &      0.33  &         3   &        3  &    0.3807  &    0.6207  &     1.692 &     0.4512 &  4.408e+09 \\\hline
\end{tabular}

{\bf Notes:}\\
$^1$ morpho-kinematics class following section 2 and Neichel et al.: 1: rotating VF and spiral morphology; 4: non-relaxed galaxies for which both kinematics and morphologies are discrepant from a rotating disk; 3: other galaxies showing either a rotating VF (and a peculiar morphology) or a complex VF and a spiral morphology. 
$^2$ M2/M1: merger mass ratio for which M1 is the mass of the observed galaxy and M2 the mass of the interloper;
$^3$ Temporal phase: before 1st pass: 1; between 1st \& 2nd pass: 2; near nuclei fusion: 3; disk rebuilding: 4; rotating disk: 5
$^4$ Merger type/orbit: DIR: 1; INC: 2; POL: 3; RET: 4; other/unknown: 5
$^5$ dTmerger: time (in Gyr) that have been spent by the galaxy in the merger
$^6$ Tmerger: total duration of the merger (in Gyr)
$^7$ f$_{gas}$ (prog): gas fraction in progenitors
$^8$ M$_{stellar}$ (prog): average stellar mass of the progenitors
\end{sidewaystable*}

\begin{acknowledgements}

We are especially indebted to Josh Barnes for making publicly his simulations of galaxy mergers on his web pages as well as letting us using the ZENO code
We are very grateful to the referee whose suggestions have considerably helped us improve the manuscript.

 \end{acknowledgements}

\end{document}